\documentclass[prb,showpacs,superscriptaddress,floatfix,twocolumn]{revtex4}
\usepackage{amsfonts}
\usepackage{stmaryrd}
\usepackage{bbm}
\usepackage{mathrsfs}
\usepackage{tipa}
\usepackage{amssymb}
\usepackage{txfonts}
\usepackage{graphicx}
\usepackage{dcolumn}
\usepackage{epstopdf}
\usepackage{array}
\usepackage{longtable}

\newcommand{\PreserveBackslash}[1]{\let\temp=\\#1\let\\=\temp}
\newcolumntype{C}[1]{>{\PreserveBackslash\centering}p{#1}}
\newcolumntype{R}[1]{>{\PreserveBackslash\raggedleft}p{#1}}
\newcolumntype{L}[1]{>{\PreserveBackslash\raggedright}p{#1}}

\begin{document}

\newcommand*{\cm}{cm$^{-1}$\,}
\newcommand*{\TlNiSe}{TlNi$_2$Se$_2$\,}


\title{Optical properties of \TlNiSe: Observation of pseudogap formation}

\author{X. B. Wang}
\author{H. P. Wang}
\affiliation{Beijing National Laboratory for Condensed Matter
Physics, Institute of Physics, Chinese Academy of Sciences,
Beijing 100190, China}

\author{Hangdong Wang}
\affiliation{Department of Physics, Zhejiang University, Hangzhou 310027, China}

\author{Minghu Fang}
\affiliation{Department of Physics, Zhejiang University, Hangzhou 310027, China}
\affiliation{Collaborative Innovation Center of Advanced Microstructures, Nanjing 210093, China}

\author{N. L. Wang}
\email{nlwang@pku.edu.cn} \affiliation{International Center for Quantum Materials, School of Physics, Peking University, Beijing 100871, China}
\affiliation{Collaborative Innovation Center of Quantum Matter, Beijing, China}

%

\begin{abstract}

Quasi-two-dimensional nickel chalcogenides \TlNiSe is a newly discovered superconductor. We have performed optical spectroscopy study on \TlNiSe single crystals over a broad frequency range at various temperatures. The overall optical reflectance spectra are similar to those observed in its isostructure $BaNi_2As_2$. Both the suppression in $R(\omega)$ and the peaklike feature in $\sigma_1(\omega)$ suggest the progressive formation of a pseudogap feature in the midinfrared range with decreasing temperatures, which might be originated from the dynamic local fluctuation of charge-density-wave (CDW) instability. We propose that the CDW instability in \TlNiSe is driven by the saddle points mechanism, due to the existence of van Hove singularity very close to the Fermi energy.

\end{abstract}

\pacs{74.70.Xa,74.70.-b,74.25.Gz,74.72.Kf}

\maketitle

\section{Introduction}

The discovery of iron-based superconductors is the most significant breakthrough in the condensed matter physics in recent years\cite{Hosono}. The iron-pnictides/chacogenides are close to magnetism and it is widely believed that the spin-fluctuation is responsible for the pairing of the superconducting electrons\cite{M1,M2}. Compared with the iron-based compounds, the nickel-based systems have much lower critical temperature, for example, LaONiP ($T_c=3 K)$\cite{LaONiP}, LaONiAs ($T_c=2.7 K)$\cite{LaONiAs}, BaNi$_2$P$_2$ ($T_c=2.4 K)$\cite{BNP}, BaNi$_2$As$_2$ ($T_c=0.7 K)$\cite{BNA1}, usually lower than 5 K even upon doping\cite{LaONiAs}. To date, no evidence for ordered or even fluctuated magnetism associated with Ni in proximity to superconductivity is found. Moreover, the majority of evidence suggests that the Ni-based systems can be understood in context of conventional electron-phonon theory\cite{T1,T2,E1,E2}.

Recently, there nickel chalcogenides KNi$_2$Se$_2$ ($T_c \simeq 0.8 K)$\cite{KNiSe}, KNi$_2$S$_2$ ($T_c \simeq 0.46 K)$\cite{KNiS}, and \TlNiSe ($T_c \simeq 3.7 K)$\cite{TlNiSeF} were reported, in which superconductivity appears to involve heavy-fermion behavior below a coherent temperature $T_{coh}$$\sim$20 K. In contrast to K$_x$Fe$_{2-y}$Se$_2$\cite{KFeSe1,KFeSe2}, \TlNiSe compound is homogeneous without Ni vacancy or phase separation\cite{TlNiSeF,KNiSe}. Considering the monovalent Tl or K in the stoichiometry composition, the effective valance of Ni in such compound is "1.5+". The mixed valency of $Ni^{1.5+}$ was proposed to induce the heavy effective band mass\cite{KNiSe,TKNiSe,KNiS}. However, very recently, angle-resolved photoemission spectroscopy (ARPES) measurement\cite{ARPES} reveals that \TlNiSe shares a universal band structure with BaFe$_2$As$_2$\cite{BFA1} and BaCo$_2$As$_2$\cite{BCA}, with a chemical potential shifted due to more $3d$ electrons in \TlNiSe (3$d^{8.5}$) than in BaFe$_2$As$_2$ ($3d^6)$ and BaCo$_2$As$_2$ ($3d^7)$. Therefore, it can be viewed as heavily electron doped TlFe$_2$Se$_2$ ($3d^{6.5})$, with reduced electronic correlations than in its cousin K$_x$Fe$_{2-y}$Se$_2$\cite{KFeSe3}. Furthermore, the camelback-shaped band at Z point gives rise to a pronounced van Hove singularity (VHS) near Fermi energy ($E_F$), which provides a natural explanation to the heavy electrons behavior inferred from the electronic heat and the upper critical field measurements\cite{TlNiSeF} in this weakly correlated system.

More specially, the neutron pair-distribution-function (PDF) analysis in a combined high-resolution synchrotron x-ray diffraction and neutron scattering study in KNi$_2$Se$_2$ reveals that the local charge density wave (CDW) state is present up to 300 K but disappears on cooling in the "heavy-fermion" state below $T_{coh}$\cite{KNiSe,KNiS}. The effect is opposite to what is typically observed. Moreover the ARPES did not observe any anomaly in the spectra upon entering the fluctuating CDW state\cite{KNiSe-ARPES}. To understand the unique behavior, it is highly desirable to investigate the charge dynamics in the whole temperature range below 300 K.

Optical spectroscopy is a powerful bulk sensitive technique with high energy resolution, which is widely used to probe the carrier dynamics and possible gap formation of an electronic system. In this work, we present a detailed optical spectroscopy study on \TlNiSe single crystals. Our optical data reveals that \TlNiSe is a good metal with a distinct plasma reflectance edge, similar to its isostructure BaNi$_2$As$_2$. The overall frequency dependent reflectivity $R(\omega)$ is gradually suppressed near 2500 \cm with decreasing temperature, and a pseudogap feature appears at any temperature below 300 K. The weak gap structure is attributed to the local CDW order or fluctuations. Our measurement indicates that the local CDW order/fluctuations are further enhanced at the lowest temperature, which is different from the neutron PDF analysis on KNi$_2$Se$_2$ in its "heavy fermion state". In Combination with the earlier ARPES results, we suggest that the fluctuating CDW instability is driven by the presence of saddle points in band structure which leads to the van Hove singularity close to the Fermi level.

\section{\label{sec:level2}Experiment}

Single crystals of \TlNiSe were grown by the self-flux method\cite{TlNiSeF}. Their structure, transport, magnetic and thermodynamic properties were already reported\cite{TlNiSeF,S.Y.Li}. The frequency-dependent reflectance spectra at different temperature were measured by Bruker IFS 113v and 80v spectrometers in the frequency range from 30 to 20 000 \cm (4 meV$\sim$2.5 eV), and then the reflectance was extended to 50 000 \cm ($\sim$6.1 eV) at room temperature with a grating-type spectrometer. An \emph{in situ} gold and aluminum overcoating technique was used to obtain the reflectivity R($\omega$). Considering the small size of our samples, the data below 100 \cm are cut off for reliability. The real part of conductivity $\sigma_1(\omega)$ is obtained by the Kramers-Kronig transformation of R($\omega$). The Hagen-Rubens relation was used for low frequency extrapolation; at high frequency side, an extrapolation method using X-ray atomic scattering functions was applied to generate the high-frequency reflectivity\cite{KKT}.

\section{\label{sec:level2}Results and discussion}

Figure 1 shows the reflectance spectra of \TlNiSe single crystals over a broad energy scale at various temperatures. The value of R($\omega$) approaches unity towards zero energy and shows an increase with decreasing temperature, indicating a good metallic behavior. By lowering the temperature, we do not see any sharp changes in the optical spectra but rather weak suppression of R($\omega$) near 2500 \cm, signaling the opening of a partial gap (pseudogap). With increasing $\omega$, R($\omega$) drops quickly near 7000 \cm, known as the screened plasma edge. The relativity high edge position reveals a high carrier density, which is similar to isostructural compound, BaNi$_2$As$_2$\cite{BNA}. There is another edge structure at about 15 000 \cm which is caused by an interband transition. The reflectivity becomes roughly temperature independent at higher frequencies. The inset of Fig. 1 shows R($\omega$) data up to 50 000 \cm in a logarithmic scale at 300 K.

\begin{figure}[b]
\includegraphics[clip,width=3.2in]{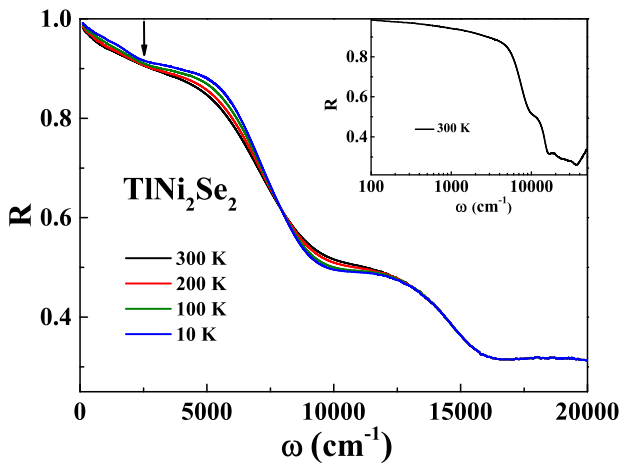}
\caption{(Color online). The temperature dependent \emph{R($\omega$)} in the frequency range from 100 to 20 000 \cm. The inset shows R($\omega$) data up to 50 000 \cm on a logarithmic scale at 300 K.}
\end{figure}

Figure 2 shows the real part of conductivity spectra below 12 000 \cm at different temperatures. The Drude-type conductivity is observed for all spectra at low frequency and several temperature-independent interband transitions could be also well resolved, as shown in the inset of Figure 2. The optical conductivity between 500 and 10 000 \cm is gradually suppressed and a broad peak-like feature becomes more and more obvious at 2800 \cm as the temperature decreases. The spectral line shape is similar to that of BaFe$_2$As$_2$\cite{BFA} across the SDW transition though the feature is much weaker. It seems to suggest that there should also be a density-wave gap opening in \TlNiSe. Furthermore, this pseudogap feature is most prominent at low temperature and can be resolved even in the 300 K data.

To quantify the temperature evolution of optical conductivity, especially the low frequency components, we use a simple Drude-Lorentz model to decompose optical conductivity spectra\cite{DL}:
\begin{equation}
\sigma_1(\omega)=\sum_i{{{\omega_{pi}^2}\over{4\pi}}{\Gamma_{Di}\over{\omega^2+\Gamma_{Di}^2}}}+\sum_j{{{S_{j}^2}\over{4\pi}}{\Gamma_{j}\omega^2\over{(\omega_{j}^2-\omega^2)^2+\omega^2\Gamma_{j}^2}}}
\label{chik}
\end{equation}
where $\omega_{pi}$ and $\Gamma_{Di}$ are the plasma frequency and the relaxation rate of each conduction band, while $\omega_j$, $\Gamma_j$ and \emph{S$_j$} are the resonance frequency, the damping and the mode strength of each Lorentz oscillator, respectively. This model includes Drude and Lorentz terms, which approximately capture the contribution by free carriers and interband transitions. Both transport\cite{TlNiSeF,S.Y.Li} and ARPES\cite{ARPES} results suggest the multiband character of \TlNiSe, therefore we applied two Drude components analysis here, similar to the other members of "122" system\cite{BFA,BNA,DL}. Even though the Drude-Lorentz fit is somewhat arbitrary and the accuracy of the derived fit parameters is uncertain, it will not modify the key outcomes of the analysis.

\begin{figure}
\includegraphics[clip,width=3.2in]{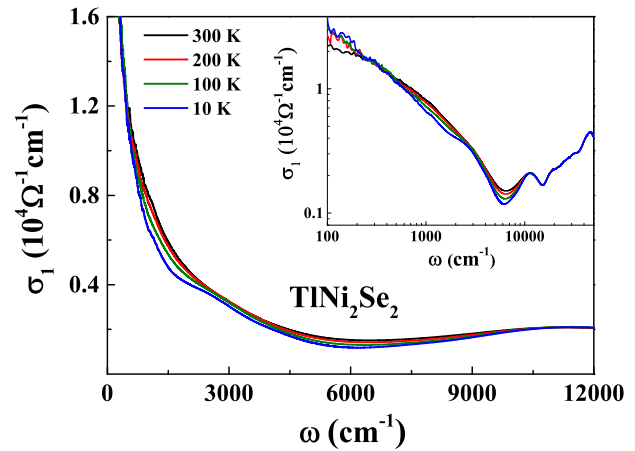}
\caption{(Color online). Frequency dependence of the optical conductivity below 12 000 \cm at different temperatures. The inset shows \emph{$\sigma_1(\omega)$} over broad frequencies up to 50 000 \cm on a logarithm scale.}
\end{figure}

Figure 3 illustrates the conductivity spectra at 300 K and 10 K together with the Drude-Lorentz fitting components at 10 K for \TlNiSe. In order to reproduce the optical conductivity below 50 000 \cm at 300 K, two Drude components, a narrow and a broad one, and four Lorentz components have to be used. Furthermore, an additional midinfrared peak is necessary near about 2800 \cm at any temperature below 300 K. We are mainly concerned about the evolution of the low energy conductivity, therefore the fitting parameters for the the two Drude components are shown in the Table 1 for different temperatures. We also show the error bars of the fitting parameters corresponding to an absolute error of 0.5\% in reflectivity. We find that the two Drude components narrow with decreasing temperature due to the metallic response. The plasma frequency $\omega_{p1}$ keeps roughly unchanged, while $\omega_{p2}$ decreases suggesting that the gapping of the Fermi surfaces mainly appears in the broad Drude component. The Lorentz 5 oscillator strength $S_5$ increases continuously with decreasing temperature, indicating the progressive formation of the absorption peaks just above the pseudogap. We use the formula $\omega_p=\sqrt{\omega_{p1}^2+\omega_{p2}^2}$ to estimate the overall plasma frequency, then we get $\omega_p \approx$ 36360 \cm at 300 K, and 32400 \cm at 10 K, respectively. Therefore, the ratio of the square of the plasma frequency at 10 K to that of 300 K is about 0.80. It is well known that $\omega _p^2 = 4\pi ne^2/m^*$, where \emph{n} is the carrier density and $m^*$ is the effective mass. If we assume that the effective mass of itinerant carriers would not change with temperature, the optical data reveals that roughly 20\% of the itinerant carriers is lost due to the pseudogap opening at the Fermi surface.

\begin{figure}[t]
\includegraphics[clip,width=3.2in]{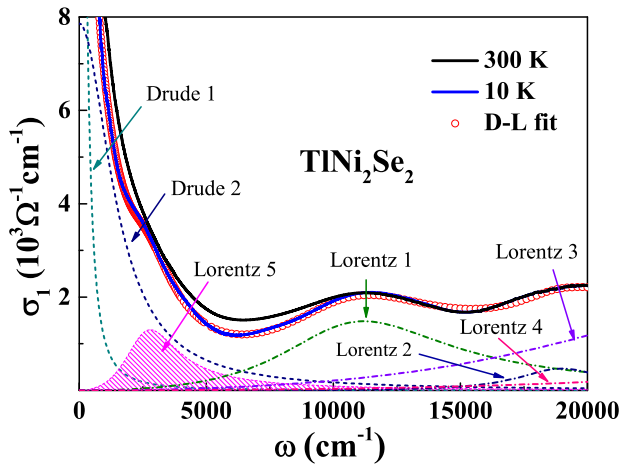}
\caption{(Color online). The experimental data of $\sigma_1(\omega)$ at 300 and 10 K with the Drude-Lorentz fits shown at the bottom.}
\end{figure}

\begin{figure}[b]
\includegraphics[clip,width=3.2in]{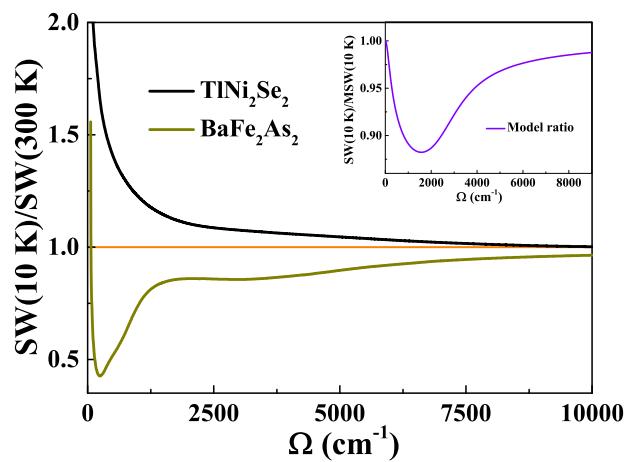}
\caption{(Color online). Ratio of the integrated spectral weight SW(10K)/SW(300K) as a function of cutoff frequency \emph{$\Omega$} in \TlNiSe and $BaFe_2As_2$. The inset shows the model ratio of the low temperature integrated SW over that of the model optical conductivity at 10 K.}
\end{figure}

\begin{table*}[htbp]
\begin{center}
\newsavebox{\tablebox}
\begin{lrbox}{\tablebox}
\begin{tabular}{*{6}{C{22mm}}c}
\hline \hline\ \\[-2ex]
T(K)&$\omega_{p1}$&$\Gamma_{D1}$&$\omega_{p2}$&$\Gamma_{D2}$&$S_5$&$\sqrt{\omega_{p1}^2+\omega_{p2}^2}$ \\[4pt]
\hline
\\[-3ex]
300 & 18500 & 390 & 31300 & 2600  & 0  & 36360\\[2pt]
200 & 18500 & 310 & 29400 & 2200  & 9500  & 34740\\[2pt]
100 & 18500 & 270 & 27700 & 1900  & 12000  & 33310\\[2pt]
10  & 18500 & 200 & 26600 & 1500  & 14100  & 32400\\[2pt]
Model& 18500 & 217 & 30100 & 1500 & 0 & 35300\\[2pt]
error bars & $\pm300$ & $\pm15$ & $\pm400$ & $\pm50$ & $\pm300$ & - \\[2pt]
\hline\hline
\end{tabular}
\end{lrbox}
\caption{The fitting parameters of two Drude components for \TlNiSe at different temperatures. $\omega_{p1}$ and $\omega_{p2}$ are the plasma frequencies of the narrow Drude term and the broad Drude term, respectively, $\Gamma_{D1}$ and $\Gamma_{D2}$ are the scattering rages of the narrow Drude term and the broad Drude term, respectively. $S_5$ is the mode strength of Lorentz 5 oscillator. The parameters of the model optical conductivity as discussed in the context is also given. The shown error bars correspond to an absolute error of 0.5\% in reflectivity. The unit of these quantities is \cm.}\scalebox{1.0}{\usebox{\tablebox}}
\end{center}
\end{table*}

To further characterize the spectral evolution, we plot the integrated spectral weight of \TlNiSe at 10 K, as shown in Figure 4, where the low-temperature integrated spectral weight has been normalized to the integrated spectral weight at room temperature. The optical spectral weight is defined as $SW=\int_0^\Omega\sigma_1(\omega)d\omega$. For the convenience of comparison, we also present the low-temperature integrated spectral weight of BaFe$_2$As$_2$\cite{BFA}. For BaFe$_2$As$_2$, the strong dip below 1000 \cm represents the formation of SDW gap in the magnetic ordered state\cite{BFA}, while the second dip near 3000 \cm reflects the unconventional SW transfer at high energy which seems to be a common feature in iron-pnictides/chacogenides\cite{SW}. Nevertheless, for \TlNiSe, the temperature-dependent spectral change are very different. The SW ratio exceeds 1 before it recovers to unity suggesting that the SW is transferred from high to low energy at low temperature. We should remark that most of the SW below 6000 \cm (the minimum of optical conductivity) would come from the intraband transitions. As the temperature decreases, the SW transformation from high to low energy is induced by the Drude components narrowing. This is the typical optical response of a metallic material. On the other hand, the pseudogap develops below 2000 \cm at low temperature. However, we note that the effects of the pseudogap formation to the SW transfer are minimal. It is the metallic response that leads to the SW transfer to low energy. In other words, the pseudogap is too weak to change the SW evolution in such a good metal. Notably, the SW of the mid-infrared peak could not fully compensate for the loss of the low-energy Drude component, and the recovery of SW extends to fairly high energy scale. Furthermore, the high energy SW transfer above 8000 \cm is almost undistinguishable in \TlNiSe. This may be related to the weaker correlation or Hund's coupling effect in \TlNiSe as compared to $BaFe_2As_2$, in agreement with the ARPES results\cite{KNiSe-ARPES,ARPES}.

In order to make the pseudogap-like suppression more clear, we also plot a model ratio of the low-temperature integrated SW to the model integrated spectral weight at 10 K. Assuming that \TlNiSe is a pure metal without the pseudogap feature, the model optical conductivity at 10 K could be obtained according to the Drude-Lorentz model with two Drude components and four Lorentz components, as the situation at 300 K . The parameters of the model optical conductivity are listed in Table 1, where $\Gamma_{D1}$ is set to 217 \cm in order to keep the same DC optical conductivity as the real data and $\omega_{p2}$ is set to 30100 \cm to conserve the total spectral weight at 10 K. We then normalize the integrated SW at low temperature SW(10 K) to the model value MSW(10 K). As shown in the inset of Fig. 4, the ratio decrease below $\approx$1700 \cm, suggesting that the SW is depleted due to opening of the pseudogap, and the suppressed SW is transferred to higher energy, leading to the gradually recovery of the SW above the pseudogap. This is a typical behavior of gap opening induced SW transformation.

Now we can summarize the main finding of our optical data: the overall optical spectra are similar to those observed in its isostructure $BaNi_2As_2$. Corresponding to the weak suppression of $R(\omega)$, the real part of the conductivity $\sigma_1(\omega)$ shows a suppression roughly below 2000 \cm, which leads to a peak above this energy. The mid-infrared peak is present at all measurement temperatures below 300 K, and becomes more and more obvious with decreasing temperature, indicating the gradual formation of pseudogap. At the same time, irrespective of the uncertainly, the overall plasma frequency decreases, implying the partial gap opening on the Fermi surface, which gives a further proof of the pseudogap formation. In particular, the SW transformation to higher energy induced by pseudogap could be clearly observed in the model ratio, where we use the metallic response at 10 K as a reference.

As we have mentioned above, for KNi$_2$Se$_2$ or KNi$_2$S$_2$, the neutron PDF analysis suggested that the local CDW fluctuations are present at high temperature but disappear upon further cooling\cite{KNiSe,KNiS}. However, our optical data of \TlNiSe single crystals reveals the progressive formation of pseudogap in the midinfrared range, especially below $T_{coh}$. To understand the origin of pseudogap in \TlNiSe crystals, there are also a few other experimental facts which should be taken into account. First, to date no evidence for magnetic transition was found on such nickel chalcogenides. Second, no static density wave, such as CDW or SDW, has been observed. Third, the anisotropy in \TlNiSe is rather small, although the compound has a layer structure. It looks like that the pseudogap in \TlNiSe might be originated from local CDW fluctuations, in agreement with earlier neutron measurement on KNi$_2$Se$_2$ and KNi$_2$S$_2$\cite{KNiSe,KNiS}. Moreover, the local CDW fluctuation was suggested to be entirely dynamic and/or spatially incoherent\cite{KNiSe,KNiS}, in contrast to the coherence CDWs observed in structurally related compounds such as \emph{2H}-typed transition metal dichalcogenides\cite{TX2}. With such CDW fluctuation, there could be a partial gap-opening in k-space on the Fermi surface, which could be related to the pseudogap feature in \TlNiSe. However, different from KNi$_2$Se$_2$ and KNi$_2$S$_2$ where the fluctuated CDW order tends to disappear upon entering the heavy electron state at very low temperature, the optical measurement on \TlNiSe indicates that the partial gap feature is further enhanced at the lowest temperature.

An earlier angle-resolved photoemission spectroscopy measurement on KNi$_2$Se$_2$ crystals suggested the quasi-two-dimensional nature of the Fermi surface and that the partial nested Fermi surface might be responsible for the local CDW fluctuations\cite{KNiSe-ARPES}. However, a more recent ARPES measurements on \TlNiSe revealed more complicated and three dimensional Fermi surface shape. Thus, a naive picture for a FS nesting scenario is not applicable here on \TlNiSe. Furthermore, a camelback-shaped band near Fermi energy at \emph{Z} point was identified, in particular, the existence of four flat part of the dispersion very close to $E_F$ would give rise to a pronounced van Hove singularity (VHS)\cite{ARPES}. On this basis, the CDW instability can be naturally understood in the frame of "saddle-point" CDW mechanism, proposed by Rice and Scott\cite{instability}. The four saddle points near \emph{Z} point in \TlNiSe gives a divergent contribution to Lindhard response function $\chi(q)$, where $q$ is a wave vector connecting two saddle points, leading to a CDW instability.

\section{\label{sec:level2}Summary}

We have performed optical spectroscopy study on \TlNiSe over a broad frequency range at various temperatures. Both the suppression in $R(\omega)$ and the peaklike feature in $\sigma_1(\omega)$ suggest the progressive formation of a pseudogap feature in the midinfrared range with decreasing temperatures, which might be originated from the dynamic local fluctuation of charge-density-wave (CDW) instability. The CDW instability in \TlNiSe may not be driven by Fermi surface nesting, but by the saddle points mechanism, due to the exitance of van Hove singularity very close to the Fermi energy.

\begin{acknowledgments}

This work is supported by the National Science Foundation of China (Grants No. 11120101003, 11327806, and 11374261), and by the Ministry of Science and Technology of China (Grant No. 2012CB821403).

\end{acknowledgments}

\end{document}